  %
  \magnification1200  \parskip 2mm plus 1mm \parindent=0pt
  
  \def\hs1{\hskip1mm} \def\h10{\hskip10mm} \def\hx{\h10\hbox}
   \def\<{\langle} \def\>{\rangle}
  \def\vs{\vskip3mm} \def\page{\vfill\eject}
  \def\Pr{{\bf Pr}}  
  \def\({\Big(} \def\){\Big)} \def\[{\Big[} \def\]{\Big]}

  \def\to{\rightarrow}

  \def\k0{{\bf k}_0}

  %


  070619
  \vs\vs\vs\vs

{\bf Quantum measurement, detection and locality}

6
  \vs\vs\h10{\bf Ian Percival$^{1,2}$ and Barry Garraway$^2$}

{\obeylines
\h10 1 Department of Physics, Queen Mary, University of London,
\h10 London E1 4NS UK
\h10 i.c.percival@qmul.ac.uk
\h10
\h10 2 Department of Physics and Astronomy, University of Sussex
\h10 Falmer, Brighton, BN1 9QH, UK
\h10 b.m.garraway@sussex.ac.uk
}

  \vs\vs\vs\vs{\bf Abstract}

  According to Bell's theorem, local realism is incompatible with quantum theory.  However, it depends on an implied assumption about quantum measurement.  We suggest that the assumption might be removed by a complete quantum analysis of the interaction between the measured system and measuring apparatus of a Bell experiment using Born's statistical interpretation.  But it is conceivable that analysis of all possible Bell experiments would allow compatibility of quantum theory and local realism.  The difficulties of theory and experiment are closely related.

  \vs\vs\vs\vs

  {\bf 1 Introduction} p2

  {\bf 2 Quantum Measurement} p3

  {\bf 3 Experiment and fair sampling} p4

  {\bf 4 Thermodynamics and detection} p5

  {\bf 5 Other evidence} p7

  {\bf 6 Conclusions} p8

  \page

  {\bf 1 Introduction}\vs

  Newton was concerned that his theory of universal gravitation allowed instantaneous transmission of gravitational force.  This problem was resolved by Einstein, whose special and general relativistic theories limit the speed of all physical influences, so that no cause can have an effect outside its forward lightcone. Classical locality follows from this classical relativistic dynamics.  However, according to Bell's theorem, idealized measurements of some entangled quantum systems produce classical correlations that are inconsistent with classical locality [1].  This is {\it weak nonlocality}.  {\it Strong nonlocality}, in which signals are sent faster than the velocity of light $c$ is still forbidden.  Quantum measurements are never ideal in practice, so Bell's theorem does not apply as it stands.  In the unlikely event that there are fundamental limits to the detection efficiencies, there could be universal locality or {\it local realism}.  

  There have been many experiments designed to demonstrate weak nonlocality, but all of them have at least one loophole.  We concentrate on the entangled photon experiment of Zeilinger, Weihs and their collaborators (ZW)[2], because it is a good experiment that has been quoted as demonstrating weak nonlocality to all intents and purposes, and giving many the impression that the experimental quest is over [3]. In an ideal experiment, both photons would be detected in every run.  In this real experiment they are not, because the quantum measurements were too far from ideal, resulting in a `detection loophole' [4].  To overcome this problem, they  assumed that their two-photon results were an unbiased sample of those from an ideal experiment.  In Section 4 we explain why this {\it fair sampling} hypothesis, which could be true for some experiments, cannot be used on its own to overcome the detection loophole.

  Santos [5.7] and one of us [6] have shown why Bell's theorem might not be enough to demonstrate weak nonlocality.  [6] and [7] use a thermodynamic analogy for illustration.  Here we go further in tracing the inadequacy of Bell's theorem to reliance on the Copenhagen theory of measurement, which provides limits on possible measurements, but no guide to how close it is possible to get to those limits.  The theorem might be completed by a quantum analysis of a Bell experiment.  Alternatively the theorem might be falsified by a comprehensive theoretical analysis of all possible Bell experiments.  

  Thus neither theory nor experiment has yet demonstrated weak nonlocality, and their difficulties are closely related.

  Like classical dynamics, quantum dynamics, or the dynamics of quantum fields, is local.  It is quantum measurements that introduce the possibility of weak nonlocality.  Here we use {\it quantum theory} to mean quantum dynamics together with a theory of quantum measurement:
  $$%
  \hbox{quantum theory = quantum dynamics + quantum measurement.} 
  $$

  Section 2 shows how Bell's theorem depends on the theory of ideal quantum measurements, and gives a recipe for its application to real experiments.  It also proposes a programme for a realistic theory of quantum measurement.  Section 3 gives a sketch of the ZW experiment, and how the fair sampling hypothesis was used to obtain a violation of the Bell-CHSH inequality.  Section 4 describes the analogy between quantum measurement and thermodynamics, showing that the violation of the inequality requires a further assumption.  Universal locality cannot be ruled out by Bell's theorem without a theoretical study of the physics of detection.  Section 5 discusses the evidence for weak nonlocality in the light of other experimental evidence, and our conclusions follow in section 6.

  \vs\vs{\bf 2 Quantum measurement}\vs

  Bell's first account of his theorem [1] refers to Dirac's presentation of the theory of observables, according to which: {\it When we make an observation we measure some dynamical variable} [8].  Unfortunately we never know exactly what dynamical variable it is that we measure.  The target dynamical variable that we intend to measure and the real dynamical variable that we actually measure are not the same. 

  Suppose we want to discriminate between zero photons and one or more photons crossing the front surface of a photodetector in a given time interval.  An ideal photodetector would click if there were one or more photons, and would not click if there were none.  In practice a good current photodetector for photons in the visible range has an efficiency not much better than 60\%, so that the detector often does not click when it should.  The efficiency is determined by the properties of a system made up of the external quantum electromagnetic field and initial quantum state of the detector.  So what is really measured is a property of an unknown state of the external field and the detector combined.

   A quantum measurement is ideal only when the target observable alone controls the classical output of the measuring apparatus.  This can never be achieved, because of the finite temperature of the apparatus, but it might be possible in some circumstances to approach arbitrarily close to this ideal, which is more than is needed to demonstrate weak nonlocality.  Without a positive answer to this question, Bell's theorem may not apply to the real world.

  So if we are to apply Bell's theorem to real systems, we need a quantum theory of the relevant parts of the detectors and their interaction with the entangled quantum system that is being measured.  This is the quantum theory of Bell experiments.  So the completion of Bell's theorem brings the theory and experiment closer together.  The detection of a photon or the measurement of its polarization are very elementary measurements, with the simplest possible observable consisting of a projector that divides the state space into two orthogonal subspaces.  Even here quantum measurements are far from ideal.  We are even further from an ideal measurement of the complete set of commuting variables of an arbitrary quantum system described in the elementary textbooks on quantum theory.

  Here we propose a programme to complete the proof of Bell's theorem by making a quantum analysis of a system designed to demonstrate weak nonlocality.  It does not use the general theory of ideal quantum measurements, but the earlier more specific statistical interpretation of Born, which introduced probability into quantum theory [9].  This is used as a theory of particle detection.  

  A complete and general theory of quantum measurement is still to be formulated.  It would include the current theory of ideal measurements as an absolute limit on what can be achieved, but would also show what is possible within these limits.

  \vs\vs{\bf 3 Experiment and fair sampling}\vs

  All the experiments designed to demonstrate weak nonlocality through the breaking of a Bell inequality have had at least one loophole.  The principal loopholes are the lightcone or locality loophole and the detection loophole.

  Figure 1 is a formalized sketch of the Zeilinger-Weihs experiment [2].  The central downconverter is pumped by a cw laser.   For each successful run it produces a two-photon state with entangled polarizations, whose frequencies sum to the laser frequency.  The polarization state is designed to have the form
  $$%
  |\chi\>= {1\over\sqrt{2}}\Big(|+\>_A|-\>_B-|-\>_A|+\>_B\Big),
  \eqno(1)$$
  where $|+\>$ and $|-\>$ are two orthogonal polarizations.  This state is independent of the choice of coordinate system for the polarizations.

  A telescope was used in the 351nm pump beam to enhance the coupling of the downconverted photons into two fibres, where they pass to Alice's and Bob's apparatus, which are similar and operate similarly.  Alice has a choice of two possible input angles $\alpha$ or $\alpha'$ for rotating the measured polarization angle, which were $0^o$ and $45^o$ for ZW.  Her photon is sent to one of two detectors, labelled $j_A=+$ and $j_A=-$ in the figure.  The first represents polarization parallel to the chosen angle, and the other perpendicular.  Bob's setup was similar.  His choice of angles $\beta$ $\beta'$ was $22.5^o$ or $67.5^o$ The outputs $j_A$ and $j_B$ are labelled $+$ or $-$ depending on which detector clicks.  All the inputs and outputs are classical events.

  In a Bell experiment with two entangled photons, whose states are measured by Alice and Bob, the lightcone loophole is failure to satisfy the relativistic conditions which ensure that no signals can pass between Alice and Bob during the course of the experiment.  In the experiments with entangled photons that satisfy these conditions, the {\it detection loophole} remains.

  Figure 2 is a simplified spacetime diagram of the ZW experiment.  DC is the downconverter.  The slopes of the lines represent speeds, where $c$ is the velocity of light in vacuo, and those with a greater slope represent the lower speed in the fibres.  $t_{\rm in}$ is the time for the start of the choice of Alice's input angle in the lab frame.  Special relativity implies that there is no time for any information about this choice to propagate to Bob by the time $t_{\rm out}$ that he completes recording his output polarization.  Similarly for Bob's input and Alice's output.  From the figure, the relativistic conditions needed to demonstrate weak nonlocality are satisfied by a wide margin.  That is, there is no lightcone loophole in this experiment.

  In an ideal experiment both photons would be detected for every run.  Failure to do so can lead to the detection loophole.  For the pair of angles $\alpha$, $\beta$, the expectation for the product $j_A j_B$ is 
  $$%
  E(\alpha, \beta) = \Pr(\alpha, \beta \to +,+)+\Pr(\alpha,\beta \to -,-)-\Pr(\alpha,\beta \to +,-)-\Pr(\alpha,\beta \to -,+) 
  \eqno(2)$$
  and similarly for the pairs of angles $\alpha,\beta'$, $\alpha',\beta$ and $\alpha',\beta'$.  The Bell-CHSH inequalities which must be satisfied if there is locality are [10]:
  $$%
  S(\alpha,\alpha',\beta,\beta') = |E(\alpha,\beta)-E(\alpha',\beta')|+|E(\alpha,\beta')+E(\alpha',\beta)|\leq 2.
  \eqno(3)$$ 
  This is for an experiment with ideal quantum measurements, in which both photons are detected for every run.  No experiment is ideal, and in the ZW experiment, as in other such experiments based on entangled photons, for the majority of runs the number detected was 0 or 1.    Gisin and Gisin [11] found a local realistic model for the experiment, so a further assumption is needed.


  To a good approximation, the ratio of the two-photon probabilities to each other was found to be the same as it would have been for an ideal experiment as predicted by the Copenhagen theory.  This is {\it fair sampling}.  So the normalized expectation for the product $j_A j_B$ for a given $\alpha,\beta$ becomes [2]
  $$%
  E_{\rm fs}(\alpha, \beta)  =
  $$
  \vskip-6mm
  $$%
  {\Pr(\alpha, \beta \to +,+)+\Pr(\alpha,\beta \to -,-)-\Pr(\alpha,\beta \to +,-)-\Pr(\alpha,\beta \to -,+)\over \Pr(\alpha, \beta \to +,+)+\Pr(\alpha,\beta \to -,-)+\Pr(\alpha,\beta \to +,-)+\Pr(\alpha,\beta \to -,+)},
  \eqno(4)$$
  and $S_{\rm fs}$ is given by equation (3) with $E_{\rm fs}$ instead of $E$.

  With this definition, the result of the ZW experiment was 
  $$%
  S_{\rm fs}(\alpha,\alpha',\beta,\beta') = 2.73\pm 0.02 >2 \hx{(30 standard deviations).}
  \eqno(5)$$

  According to the classical theory of locality given above, all output events $(j_A,j_B)$ must be included in the analysis, and for ZW this includes the detection of one photon by Alice or by Bob, which was much more probable than two-photon detection.  When these events are included, the CHSH inequality (3) no longer applies.  But according to the fair sampling hypothesis, {\it if} the efficiency of detection was increased sufficiently, locality would follow.  This is a big {\it IF}.  For such a fundamental question as the strict locality of physical laws, we should not accept this assumption if there is an alternative consistent with current experiments, as discussed in the next section.

  \vs\vs{\bf 4 Thermodynamics and detection}\vs

  The Copenhagen theory of quantum measurement defines a limit on what can be achieved in determining the state of a quantum system: the values of a complete set of commuting observables.  It describes what could be achieved if it were possible to carry out ideal measurements, not what has been achieved in current laboratory experiments.  It would not be possible to carry out a Bell experiment for entangled neutrinos, because the weak interaction of the neutrino with the charged particles that make up ordinary matter prevent even the reliable detection of a single neutrino.  The interaction of a photon with charged particles is stronger than for a neutrino, but the practical detection problem is still significant for the entangled photons near optical wavelengths that are used in experiments on locality.  The question is whether or not this practical problem has its origin in any fundamental limit on detection efficiency.

  Clearly the constraints on measurement of photons are much less severe than for neutrinos.  Nevertheless, to demonstrate weak nonlocality theoretically, measurement constraints which ensure locality have to be ruled out.  Otherwise the fair sampling hypothesis cannot be applied.  It might not even be possible in any set-up, in which case our world would be strictly local.  Bell wrote [12]:  `\dots it is difficult for me to believe that quantum mechanics, working very well for inefficient practical set-ups, will nevertheless fail badly with improvements in counter efficiency and other factors ...'   This assumes that these improvements {\it can} be made.  The danger in this assumption can be seen through a thermodynamical analogy [6,7]: The first law of thermodynamics relates heat energy and work.  The second law is consistent with the first, but puts a constraint on the efficiency of conversion.  Conceivably there could be a similar second law that puts a constraint on the efficiency of quantum measurement and which is consistent with the Copenhagen rules.

  In practice, Bell measurements are based on local detection.  The classical output events for Bell experiments are not ideal measurements, but physical detections using physical detectors, whose limitations contribute to the detection loophole.  Using fair sampling to extrapolate from inefficient detection to efficient detection, as in the previous section, assumes that the efficient detection can be achieved.  Experimentally, weak nonlocality can only be demonstrated by an experiment without loopholes.

  As for theory, the Copenhagen theory of ideal quantum measurements is inadequate.   Born's earlier statistical theory is a theory of particle detection [9].  When applied with a reasonable choice of boundary between the quantum and classical, it does not, like the later Copenhagen interpretation of quantum measurement, assume anything equivalent to ideal measurements.  It is always to be preferred over Copenhagen where the deviation from ideal may be significant, as in Bell experiments.   

  Completing Bell's theorem brings theory and experiment closer together.  A single successful Bell experiment without loopholes would confirm nonlocality.  Experimentally, anything less needs further questionable assumptions.  The original form of Bell's theorem is not enough to demonstrate nonlocality either, because it depends on the assumption that it is possible to approach sufficiently close to ideal quantum measurements.  Of course, this would be demonstrated by a successful Bell experiment with no loopholes, but is it possible to complete Bell's theorem theoretically?  Clearly the Copenhagen theory of measurement can't be used, but a form of Born's theory can.  

What is needed is a complete quantum theory of a Bell {\it thought} experiment, including a theory of all its parts.  For a two-photon entanglement experiment like ZW, this requires the detailed quantum theory of the interaction of a photodetector with an external quantized electromagnetic field, including the final electron avalanche up to the stage where it satisfies classical dynamics.  This requires the solution of the Schr\"odinger equation for all parts of the system, including the apparatus and the photons, which affect the causal links between classical inputs and outputs.  Here the inputs are at the source S and the classical subsystems that control the direction of polarization measurements at A and B.  The outputs are the classical signals at the detectors.   If this analysis showed that a Bell inequality could be broken, then the result of Bell's theorem would follow.  Such a complete analysis is not easy.

  {\it Any} loophole-free Bell thought experiment would do, even if the corresponding laboratory experiment was prohibitively expensive, or depended on planned future technology with a sound theoretical basis.  If it could be shown that no such thought experiment was possible, then Bell's theorem would be false.

  Theoretical analysis is needed for the design of proposed Bell experiments.  Here we suggest that the analysis has an importance in its own right for the completion of Bell's theorem, or even conceivably for its rejection.  Suppose that the theory of an experiment supported Bell's theorem.  This would either show that weak nonlocality followed from quantum dynamics and Born's theory of detection, or alternatively that locality implies that Born's theory cannot be applied to entangled quantum systems, since the evidence for quantum dynamics is so strong.  Nevertheless, it would still be worth doing the Bell experiment, since no-one has yet either demonstrated weak nonlocality or found fault with Born's theory.  Suppose instead that the theory showed that a planned experiment could not break a Bell inequality.  This would suggest that the experiment should be modified, or a different experiment tried.  It would not demonstrate universal locality.  That would require a much more general proof that for any conceivable set of measurements on entangled quantum systems, no Bell inequality could be broken: a formidable task, which may well be impossible.

  It should be mentioned here that when Bohm illustrated the theory of measurement for the de Broglie pilot-wave theory [13], he was forced to consider the dynamics of the interaction between the measured quantum system and the measuring apparatus, just as we have suggested for Bell's theorem.  But Bohm only needed to consider a model interaction, whereas for our purposes a detailed examination of an experiment is necessary.

\vs\vs  {\bf 5 Other evidence}\vs

Using a sophisticated downconverter design, Altepeter, Jeffrey and Kwiat [14] obtained a violation of CHSH by $1239\sigma$, assuming fair sampling.  They reported a detection efficiency of 30\%, a significant improvement over ZW, but still not enough to overcome the detection loophole.  Loopholes were not mentioned.
 
  The argument for nonlocality is not based only on photon entanglement experiments like ZW.  There are also experiments with massive particles that do not satisfy the lightcone condition, but which overcome the detection loophole.  Prominent among these is the experiment of Wineland, Rowe and their collaborators [15], who overcame the detection loophole with entangled ions, but their experiment failed to satisfy the relativistic conditions necessary to overcome the lightcone loophole by a factor of more than $10^{10}$, a formidable factor to overcome.  Further the ions were so close that two of the outputs from Alice and Bob's ions were indistinguishable.  As they pointed out, this does not affect the CHSH inequality.  But it does affect the answer to the question: does this experiment satisfy all the conditions necessary to demonstrate nonlocality apart from the lightcone condition?  One necessary condition is that Alice's and Bob's outputs are distinguishable, so the answer is no.  Nevertheless, the experiment shows that there are circumstances in which the detection loophole can be overcome.

  Fry, Walther and Li [16,17] have published a theoretical analysis of a proposed experiment based on entangled Hg atoms produced by dissociation of dimers, whereas Garc\'ia-Patr\'on et. al. [18] have proposed using squeezed entangled states and homodyne detection.  We suggest that any thorough theoretical analysis of a proposed Bell experiment is worth publishing, even if the resources for the experiment are not available, because such analysis is a step towards completing Bell's theorem.

  Since the relativistic lightcone condition makes detection more difficult, it is worth asking whether it makes physical sense for there to be universal locality with either the lightcone or detection loopholes present in all possible experiments.  Perhaps the Bell inequalities could be broken for nonrelativistic experiments that fail to satisfy the lightcone condition, but not otherwise.  In this, the photon experiments provide some help, because it is usual to use the same apparatus for small distances between Alice and Bob, breaking the lightcone condition, and for the large distances necessary to satisfy the condition, although the comparison is not generally published [19].  The unpublished results of which we are aware are consistent with continuity through the light cone.  To our knowledge, no experiment has been specifically designed to investigate any changes in a Bell inequality on passing through the light cone.

  \vs\vs{\bf 6 Conclusions}\vs

  According to Bell's theorem, local realism is inconsistent with the Copenhagen theory of quantum measurement.  The theorem is incomplete, because of over-reliance on the Copenhagen theory.  The theorem might be completed by a complete theoretical analysis of a single Bell thought experiment.

As an example of the difference between a real Bell experiment and a Bell thought experiment, consider the thought that Bell experimenters had the resources of big science.  ZW and others used optical fibres to transmit the photons from the source to A and B.  The resultant losses due to injection are very significant.  Why not focus the photons from the source onto A and B instead?  Because of diffraction and absorption in air.  The diffraction could be made negligible by using a sufficiently large adapted astronomical telescope.  The absorption in air by putting the telescope in a vacuum, using the resources of CERN.

 Conceivably the theoretical analysis could show instead that no adequate measurements are possible, thus allowing local realism.  This would follow if there were a fundamental limit on detection efficiency in all Bell experiments.  A complete quantum analysis of a Bell experiment is needed for Bell's theorem, and is also needed for experimental design, showing a convergence of theory and experiment.  Any thorough theoretical analysis of a proposed Bell experiment is worth publishing, even if the experiment is too expensive to perform, because such analysis helps to complete Bell's theorem.  It is probable that detection efficiency can be improved enough to ensure nonlocality, but this has neither been proved using an adequate theory, nor demonstrated experimentally without loopholes.  Consequently fair sampling cannot be used to close the detection loophole.

 The Copenhagen rules of quantum measurement provide limits to quantum measurement, but do not show how close we can get to these limits.  Born's statistical interpretation can be considered as a theory of quantum detection.  It can then be used to study the approach to those limits, which is needed if we are to have a complete quantum measurement theory.  Bell's theorem could be completed by a full analysis of a single Bell thought experiment using quantum dynamics and Born's theory, including the physics of each part of the apparatus from the production of the entangled system to the final detection of the particles.  This suggests that more experimenters should publish the theory of their experiments, and that theoreticians who are interested in completing Bell's theorem might be more closely involved in experimental design.

The Copenhagen theory of measurement is incomplete, so that Bell's theorem is also incomplete.  For similar reasons, the fair sampling assumption of Bell experiments, even if true, does not overcome the detection loophole.  So it is still conceivable, though improbable, that all physical laws are local.  

\vs{\bf Acknowledgements}

We are grateful to Gernot Alber, Todd Brun, Jessica Cheung, Nicolas Gisin, Serge Haroche, Chris Chunnilul, Tim Spiller, Walter Strunz, Thomas Walther and our colleagues at QMUL and Sussex for very helpful communications. ICP thanks the Leverhulme Trust for financial support.


\vs{\bf References} 

[1] Bell J S 1988 {\it Speakable and Unspeakable in Quantum Mechanics} Cambridge University Press, Cambridge, U.K., p14, p19.  Also 1964 {\it Physics} {\bf 1} 195

[2] Weihs G, Jennewein T, Simon C, Weinfurter H and Zeilinger A 1998    
{\it Phys. Rev. Lett.} {\bf 81} 5039 

[3]  Aspect A 1999 {\it Nature} {\bf 398} 189

[4] Pearle P 1970 {\it Phys. Rev. D} {\bf 2} 1418 

[5] Santos E 1992 {\it Phys. Rev. A} {\bf 46} 3646 

[6] Percival I 2001 {\it Phys. Lett. A} {\bf 279} 105 

[7] Santos E 2005 {\it Studies in the History and Philosophy of Modern Physics} {\bf 36B} 544
[8] Dirac P A M 1947 {\it The Principles of Quantum Mechanics} Oxford, Clarendon Press, 3rd edn., p34, p37

[9] Born M 1926 in Wheeler J A and Zurek W H (eds) 1983 {\it Quantum Theory and Measurement} Princeton, New Jersey, University Press, p52 

[10] Ref [1] p156

[11] Gisin N and Gisin B 1999 {\it Phys. Lett. A} {\bf 260} 323  

[12] Ref [1], p109

[13] Bohm D 1952 {\it Phys. Rev} {\bf 85} 180 

[14] Altepeter J, Jeffrey E and Kwiat P 2005  {\it Optics Express} {\bf 13} 8951

[15] Rowe M A, Kielpinsky D, Meyer V, Sackett C A, Itano W M, Monroe C and Wineland D J 2001 {\bf 409} 791

[16] Fry E S, Walther T and Li S 1995 {\it Phys. Rev. A} {\bf 52} 4381 

[17] Fry E S and Walther T 1997  {\it Experimental Metaphysics}  (Kluwer, Dordrecht, Netherlands,  eds Cohen R S, Horne M and Stachel J) p61    

[18] Garc\'ia-Patr\'on R, Fiur\'a\v sek, Cerf N J, Wenger J, Tualle-Brouri R and Grangier Ph 2004 {\it Phys. Rev. Lett.} {\bf 93}, 130409 

[19] Private Communication from Nicolas Gisin
\page

FIGURE CAPTIONS.
\vs\vs

FIG 1

Sketch of apparatus for the ZW experiment.  The downconverter is labelled dc.
Alice rotates the angle of polarization of her photon so that polarization
parallel to the angle $\alpha$ is detected in detector $+$, and perpendicular
in detector $-$.  Similarly for Bob with angle $\beta$.
\vs

FIG 2

Rough spacetime diagram of the ZW experiment.  Light propagates with speed $u$ in the optical fibres.  The time intervals $t_i,t_j$ in the laboratory frame between Alice's and Bob's inputs and outputs are represented by the short vertical lines.  They lie entirely outside each-other's light cones.

\end